\documentclass[prl,amsmath,amssymb,superscriptaddress, twocolumn ]{revtex4-1}
\usepackage{graphicx}
\usepackage{epsfig}
\usepackage{dcolumn}
\usepackage{bm}
\usepackage{rotating}
\usepackage{multirow}
\usepackage[table]{xcolor}
\usepackage[english]{babel}
\newcommand{\rr}{\mathbf{r}}
\newcommand{\rrp}{\mathbf{r'}}

\begin{document}
\widetext 

\title{Bond breaking and bond formation: 
how electron correlation is captured in many-body perturbation theory and density-functional theory}

\author{Fabio Caruso}
\affiliation{Fritz-Haber-Institut der Max-Planck-Gesellschaft, Faradayweg 4-6, D-14195 Berlin, Germany}
\author{Daniel R. Rohr}
\affiliation{Department of Chemistry, Rice University, Houston, Texas 77005, USA}
\affiliation{Fritz-Haber-Institut der Max-Planck-Gesellschaft, Faradayweg 4-6, D-14195 Berlin, Germany}
\author{Maria Hellgren}
\affiliation{SISSA, International School for Advanced Studies, via Bonomea 265, 34136 Trieste, Italy}
\author{Xinguo Ren}
\author{Patrick Rinke}
\affiliation{Fritz-Haber-Institut der Max-Planck-Gesellschaft, Faradayweg 4-6, D-14195 Berlin, Germany}
\author{Angel Rubio}
\affiliation{Nano-Bio Spectroscopy group and ETSF Scientific Development Centre,
Universidad del Pa\'is Vasco, CFM CSIC-UPV/EHU-MPC and DIPC, Av.\ Tolosa 72, E-20018 Donostia, Spain}
\affiliation{Fritz-Haber-Institut der Max-Planck-Gesellschaft, Faradayweg 4-6, D-14195 Berlin, Germany}
\affiliation{European Theoretical Spectroscopy Facility}
\author{Matthias Scheffler}
\affiliation{Fritz-Haber-Institut der Max-Planck-Gesellschaft, Faradayweg 4-6, D-14195 Berlin, Germany}
\date{\today}

\date{\today}
\begin{abstract}
For the paradigmatic case of H$_2$-dissociation we compare state-of-the-art many-body perturbation theory (MBPT) in the $GW$ approximation and density-functional theory (DFT) in the exact-exchange plus random-phase approximation for the correlation energy (EX+cRPA). 
For an unbiased comparison and to prevent spurious starting point effects both approaches are iterated to {\it full} self-consistency (i.e. sc-RPA and sc-$GW$).
The exchange-correlation diagrams in both approaches are topologically identical, but in sc-RPA they are evaluated with non-interacting and in sc-$GW$ with interacting Green functions. This has a profound consequence for the dissociation region, where sc-RPA is superior to sc-$GW$. We argue that for a given diagrammatic expansion, sc-RPA  outperforms sc-$GW$ when it comes to bond-breaking. We attribute this to the difference in the correlation energy rather than the treatment of the kinetic energy.
\end{abstract}

\maketitle
First-principles electronic-structure calculations have become indispensable in many fields of science, because they yield atomistic insight and are complementary to purely experimental studies. Since the full many-body problem of interacting electrons and nuclei is intractable for all but the simplest systems, different strategies for approximate approaches have been developed over the years. The most prominent are density-functional theory (DFT) \cite{hohenbergkohn,kohnsham1965,Dreizler}, many-body perturbation theory (MBPT)~\cite{Szabo/Ostlund:1989,fetter,Hedin1965}, coupled-cluster theory \cite{RevModPhys.79.291} and quantum Monte Carlo methods \cite{RevModPhys.73.33}. 
Each approach has its strengths and weaknesses in terms of accuracy, applicability, and computational efficiency and no consensus has been reached regarding the optimal approach for current and future challenges in electronic-structure theory. In this work we address the difference between DFT and MBPT for the total energy and ask the questions: Given a fixed set of diagrams for the electron-electron interaction, will the DFT and the MBPT framework give the same result? And if not, which one is better? To answer these  questions we consider the paradigmatic case of H$_2$ dissociation. Other diatomic molecules are presented in the Supplemental Material.

\begin{figure} 
\includegraphics[width=0.48\textwidth]{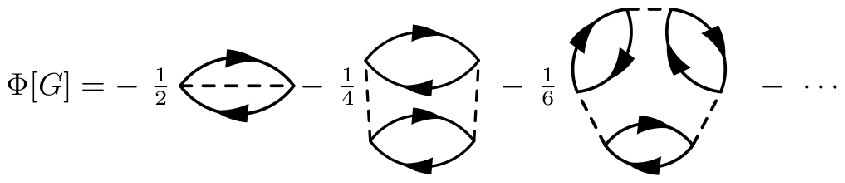}
\caption{\label{fig:phi} $\Phi$ functional for RPA and $GW$ correlation energies (Eq.~\ref{eq:phifunc}). The arrowed lines
 correspond to the interacting Green function $G$ in $GW$, and
 the KS Green function $G_s$ in RPA. 
 Dashed lines denote the bare Coulomb interaction, 
 and the minus sign of the prefactor comes from the rules for evaluating Feynman diagrams \cite{baymkadanoff,luttingerward1960}.
 }
\end{figure}

In the past, DFT and MBPT have been compared directly in the exchange-only case \cite{Gorling/Ernzerhof/OEP/HF}. In MBPT this corresponds to the Hartree-Fock approach, whereas in DFT a multiplicative Kohn-Sham (KS) potential is constructed by means of the optimized effective potential approach (OEP) \cite{review_OEP}. 
As we will demonstrate in this Letter, the comparison between DFT and MBPT can be extended to encompass correlation using exact-exchange plus correlation in the random-phase approximation to DFT (EX+cRPA), \cite{Bohm/Pines:1953,Gell-Mann/Brueckner:1957,Langreth/Perdew:1977,RPAreview} referred to as RPA in the following, and the $GW$ approach to MBPT \cite{Hedin1965,thegwmethod}.  The exchange-correlation diagrams in both approaches are topologically identical (see Fig.~\ref{fig:phi}), but in RPA they are evaluated with a non-interacting KS and in $GW$ with an interacting Green function. 
To illustrate the impact of these differences 
we consider the bond-breaking/formation regimes in the binding curves of H$_2$.
To avoid starting point effects both approaches are iterated to self-consistency, which we denote as sc-RPA and sc-$GW$.
The extension of this study to higher order correlation diagrams is, in principle, possible and will be pursued in 
future work. Here, we focus on sc-RPA and sc-$GW$ as they provide the simplest (and currently only computationally 
tractable) way for more complex systems to compare density-functional and many-body theory.

\begin{figure*}
\includegraphics[width=0.99\textwidth]{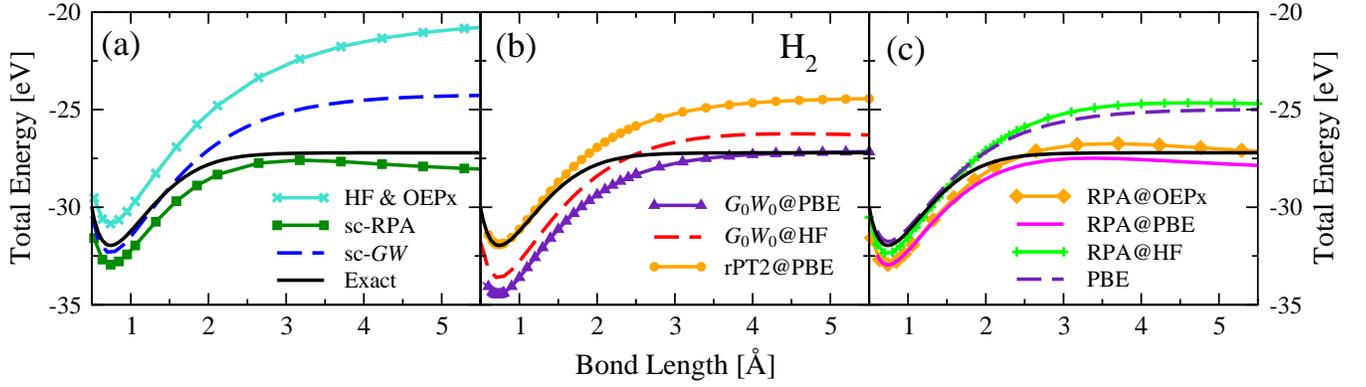}
\caption{\label{fig:h2} (Color online) Total energy (eV) of the H$_2$ molecule as  a function of bond length (\AA). Different  flavors of $GW$ and RPA are shown compared to PBE, rPT2 and accurate full configuration interaction calculations  taken from 
Ref.~\cite{h2ci1993}. Hartree-Fock (HF) and exact-exchange OEP (OEPx) are identical for H$_2$ and are included for comparison. All calculations were performed using a Gaussian cc-pVQZ \cite{gaussianbasis1989} basis set. 
}
\end{figure*}

Let us start with the ground-state total-energy expression for an interacting electron system obtained with
the adiabatic-connection (AC) technique (see e.g. Ref.~\cite{RPAreview}): 
 \begin{align}
       E =& E_0 - \frac{1}{2} \int_0^1 d\lambda \int d\rr d\rrp v({\bf r},{\bf r'})  \times \nonumber \\
         &\left[ \int_0^\infty \frac{d\omega}{\pi} \chi_\lambda(\rr,\rrp;i\omega) +  
                n(\rr)\delta(\rr-\rrp)
                     \label{Eq:E_ACFD}   \right] \\
       =& E_0 + 
        \int_0^1 \frac{d\lambda}{\lambda} \int_{0}^\infty \frac{d\omega}{2\pi}
          \text{Tr}\left[\Sigma_\lambda(i\omega) G_\lambda (i\omega) \right] \, .
        \label{Eq:E_Sigma-G}
 \end{align}
Here $v({\bf r},{\bf r'})$ is the Coulomb interaction, $\text{Tr}\left[AB\right]$ denotes $\int d\rr d\rrp A(\rr,\rrp)B(\rrp,\rr)$ and $ E_0 = T_\text{s} + E_{\rm H} + E_{\rm ext}$.  
$T_\text{s}$ is the kinetic energy of the KS independent-particle system,  
$E_{\rm H}$ the Hartree and $E_{\rm ext}$ the external energy.
Along the AC path (i.e. at each value of $\lambda$), 
the electron density $n(\rr)$ 
is assumed to be fixed at its physical value, and $\chi_\lambda$, $G_\lambda$, and $\Sigma_\lambda$ are 
the polarizability, the single-particle (time-ordered) Green function, and the self-energy, respectively. 
In the following we adopt the notation $G_s\equiv G_{\lambda=0}$ for the non-interacting KS Green function and 
$G\equiv G_{\lambda=1}$ for the fully interacting one.  

The RPA for the total energy can be most conveniently
introduced in Eq.~\ref{Eq:E_ACFD} through the approximation $\chi_\lambda=\chi_s(1-\lambda v \chi_s)^{-1}$,
where $\chi_s=\chi_{\lambda=0}=-iG_sG_s$.
Within this approximation, 
the integrand in Eq.~\ref{Eq:E_ACFD} is assumed to depend on $\lambda$
only through the scaled Coulomb interaction $\lambda v$. 
Alternatively the RPA total energy can also be obtained 
through Eq.~\ref{Eq:E_Sigma-G} by introducing the $GW$ 
approximation for the proper self-energy \cite{PhysRevA.68.032507,RPAreview}: 
  \begin{equation}
   \Sigma_\lambda^{GW}(\omega) = 
  \int \frac{d\omega'}{2\pi} G_{\lambda}(\omega+\omega')W_\lambda(\omega')e^{i\omega'\eta},
  \label{Eq:GW_approx}
  \end{equation}
where $W_\lambda [G_\lambda]= \lambda v(1+i\lambda vG_\lambda G_\lambda)^{-1}$ and $\eta$ is a positive infinitesimal. 
The RPA total energy is retrieved by omitting
the $\lambda$-dependence of $G_{\lambda}$, i.e. replacing $G_{\lambda}$ by the KS non-interacting 
Green function $G_s=G_{\lambda=0}$ and $W_\lambda$ by $W_s \equiv W_\lambda [G_s]$.
Either way, the $\lambda$ integration in Eqs.~\ref{Eq:E_ACFD} or \ref{Eq:E_Sigma-G} 
can now be carried out, yielding the sum of the exact-exchange energy $E_\text{x}$ and the RPA 
correlation energy $E_\text{c}^\text{RPA}$, where:
 \begin{align}
 E_{\rm c}^\mathrm{RPA} 
   & =\int_0^\infty \frac{d\omega}{2\pi} \text{Tr} \left[ \text{ln}(1-\chi_s (i\omega) v) + 
      \chi_s (i\omega) v \right].
   \label{Eq:E_c_rpa}
 \end{align}
Combining Eqs.~\ref{Eq:E_ACFD} and \ref{Eq:E_c_rpa} allows us to express the RPA total energy functional as:
 \begin{align}
     E^\text{RPA}[G_s]
   = & T_s + E_\text{ext} + E_\text{H} +E_\text{x} + E_\text{c}^\text{RPA}\, .
   \label{Eq:E_Klein-RPA}
 \end{align}

We now come to the differences in the evaluation of the total energy in the context of
KS-DFT and MBPT.
%
\begin{table*}[hbtpd] 
\begin{ruledtabular}
\caption{Total energies (in eV) of H$_2$ in the equilibrium geometry and deviation ($\Delta$) from the exact reference.  }
\begin{tabular*} 
           {\textwidth} {@{}%
          c@{\extracolsep{\fill}}%
          d@{\extracolsep{\fill}}%
          d@{\extracolsep{\fill}}%
          d@{\extracolsep{\fill}}%
          d@{\extracolsep{\fill}}%
          d@{\extracolsep{\fill}}%
          d@{\extracolsep{\fill}}%
          d@{\extracolsep{\fill}}%
          d@{\extracolsep{\fill}}%
          d@{\extracolsep{\fill}}%
          d@{\extracolsep{\fill}}}
\multicolumn{1}{c}{\text{ }} & 
\multicolumn{1}{c}{\text{Exact \cite{h2ci1993}}} & 
\multicolumn{1}{c}{\text{sc-$GW$}} & 
\multicolumn{1}{c}{\text{sc-RPA}} & 
\multicolumn{1}{c}{\text{$G_0W_0$@HF}} & 
\multicolumn{1}{c}{\text{$G_0W_0$@PBE}} & 
\multicolumn{1}{c}{\text{RPA@HF}} &
\multicolumn{1}{c}{\text{RPA@OEPx}} & 
\multicolumn{1}{c}{\text{RPA@PBE}} & 
\multicolumn{1}{c}{\text{rPT2@PBE}} & 
\multicolumn{1}{c}{\text{HF\&OEPx}} \\
\hline
\,\,\,\,$E_{\rm tot}$ & -31.97  & -32.33 & -32.95 & -33.58  & -34.48 & -32.38 & -32.95 & -32.95 & -31.92 & -30.88 \label{tab:tab} \\
\,\,\,\,$\Delta$ & & 0.36 & 0.98 & 1.61 & 2.51 & 0.41 & 0.98 & 0.98 & -0.05 & -1.09 
\end{tabular*}                                                
 \end{ruledtabular}
\end{table*}
%
In MBPT, the  Green function $G_\lambda$ represents an interacting electron system for $\lambda\neq0$, 
and has to satisfy the Dyson equation: 
   \begin{equation}
     G_\lambda^{-1} = G_s^{-1} - \Sigma_\lambda[G_\lambda] - v^\lambda_\text{ext}+ v_\text{ext} +
         (1-\lambda)v_\text{H} +  v_\text{xc}  \,
     \label{Eq:Dyson}
  \end{equation}
with $v^\lambda_\text{ext}$ being the external potential of the $\lambda$-dependent system (chosen to keep the 
density fixed), and $v_\text{xc}$ the exchange-correlation potential of the KS non-interacting particle reference system. 
Making use of Eqs.~\ref{Eq:GW_approx} and \ref{Eq:Dyson}, the $\lambda$-integration
in Eq.~\ref{Eq:E_Sigma-G} can be carried out and one arrives at the following expression for the total energy
 \begin{align}
     E=  &-E_\text{H}[G] + \Phi[G] - \frac{1}{2\pi} \int_{-\infty}^\infty d\omega \times
       \nonumber      \\
    &\text{Tr}\left[(G_s^{-1}(i\omega) + v_\text{xc})G(i\omega) - 1 + \text{ln}(G^{-1}(i\omega)) \right] \,  .
       \label{Eq:E_Klein}
 \end{align}
Details for the derivation of 
Eq.~\ref{Eq:E_Klein} can be found in the supplemental material \cite{supple_mater}. 
In Eq.~\ref{Eq:E_Klein}, 
the functional $\Phi[G]$ is defined as \cite{baymkadanoff,luttingerward1960}
 \begin{equation}\label{eq:phifunc}
   \Phi[G] = \frac{1}{2\pi}\int_{-\infty}^\infty d\omega 
         \sum_{n=1}^\infty \frac{1}{2n} \text{Tr}\left[\Sigma^{(n)}(i\omega)G(i\omega)\right]\, ,
 \end{equation}
where $\Sigma^{(n)}$ is the sum of all self-energy diagrams that contain $n$ explicit Coulomb 
interaction lines. 
We note that since
$\Sigma = \delta \Phi / \delta G$
, an approximation for $\Phi$ 
directly translates into a corresponding approximation for $\Sigma$. The diagrammatic representation of
$\Phi$ in the $GW$ approximation is illustrated in Fig.~\ref{fig:phi}.

In the KS framework, the sc-RPA
total energy is obtained by requiring 
$G_s$ in Eq.~\ref{Eq:E_Klein-RPA} to satisfy the Dyson equation 
$G_s(i\omega)=(i\omega+\nabla^2/2-v_\text{ext}-v_\text{H}-v_\text{xc}^\text{RPA})^{-1}$, where 
$v_\text{xc}^\text{RPA}$ is determined by the optimized effective potential equation (also known as the linearized 
Sham-Schl\"{u}ter equation) \cite{ShamSchluter,Casida:1995}:
  \begin{equation}
     G_s(\Sigma^{GW}[G_s] - v_\text{xc}^\text{RPA})G_s = 0\, .
    \label{Eq:OEP}
  \end{equation}
Alternatively, the sc-RPA energy can be obtained by minimizing $E^\text{RPA}$ in Eq.~\ref{Eq:E_Klein-RPA} 
with respect to the non-interacting input KS Green functions $G_s$.

Regarding the energy expression of Eq.~\ref{Eq:E_Klein} as a functional of $G$ yields the well-known Klein 
functional \cite{klein1961}. This functional is stationary (i.e., $\delta E[G]/\delta G = 0$) at the 
self-consistent $G$ of the Dyson equation \cite{klein1961}.
It has further been shown  \cite{Casida:1995,dahlenleeuwen2006} that evaluating the Klein functional 
(using the $GW$ approximation for $\Phi$) with the KS
reference Green function $G_s$ one obtains the RPA total energy in Eq.~\ref{Eq:E_Klein-RPA} (see also the supplemental material 
\cite{supple_mater}).
This offers a second way to look at the difference between sc-RPA and sc-$GW$: 
the sc-RPA energy corresponds to a mininum of the Klein functional within a variational subspace of non-interacing
KS Green functions, whereas the sc-$GW$ total energy corresponds to a stationary point of the Klein functional
in a larger variational space including both noninteracting and interacting Green functions. 
However, we emphasize that this stationary point is not necessarily a minimum \cite{klein1961,luttingerward1960}. In practical calculations, 
the sc-$GW$ total energy is actually above the sc-RPA energy as we will show in this Letter.

For a quantitative comparison between sc-$GW$ and sc-RPA, 
we choose the Galitskii-Migdal (GM) formula \cite{galitskiimigdal} 
for the computation of the sc-$GW$ total energy. 
At self-consistency, the GM formula is coincides with the Klein functional (Eq.~\ref{Eq:E_Klein}) 
 -- as for instance discussed in Refs.~\cite{baym,dahlenleeuwen2005,dahlenleeuwen2006}. 
The GM formula can be expressed as \cite{caruso/prb/2012}:
\begin{align}
E^{GW}[G] = T+E_{\rm ext}+E_{\rm H}+E_{\rm x}+ E_{\rm c}^{GW}\, ,
\label{eq:gwtot}
\end{align}
where all terms on the right hand side of Eq.~\ref{eq:gwtot} are regarded as
functionals of the Green function $G$. $T$ is the kinetic energy 
of the interacting system and $E_{\rm c}^{GW}$ the so-called $GW$ 
correlation energy defined as 
\begin{align}
\label{eq:gwcorr}
 E_{\rm c}^{GW}[G]=\int_0^\infty \frac{d\omega}{2\pi} {\rm Tr}\{G(i\omega) \Sigma_{\rm c}^{GW}[G](i\omega)\} \, .
\end{align}
Here $\Sigma_{\rm c}^{GW}$ is the correlation part of the $GW$ self-energy. 
The evaluation of Eq.~\ref{eq:gwtot} with a HF (PBE) Green function is referred to as 
$G_0W_0$@HF ($G_0W_0$@PBE) total energy.

An inspection of Eq.~\ref{Eq:E_c_rpa} and \ref{eq:gwcorr} reveals 
that the difference in the sc-$GW$ correlation energy and 
the sc-RPA correlation energy is twofold. First, the sc-$GW$ 
expression is evaluated with an interacting Green function as 
opposed to a Kohn-Sham  one in sc-RPA. Second, the kinetic correlation energy 
-- i.e., the difference between the full kinetic energy and that of  
the non-interacting KS system -- 
are included in Eq.~\ref{Eq:E_c_rpa} through the coupling
constant integration, whereas in sc-$GW$ the correlation term is purely Coulombic.
To facilitate a term-by-term comparison between sc-$GW$ and sc-RPA total energies, we separate $E_\text{c}^\text{RPA}$ into  the Coulomb
correlation energy $U_\text{c}^\text{RPA}$ and the kinetic correlation energy $T_\text{c}^\text{RPA}$:
\begin{align}
  U_{\rm c}^{\rm RPA}&= - \int_0^\infty \frac{d\omega}{2\pi} \text{Tr} 
  \left[ \sum_{n=2}^{\infty}{\left(\chi_s(i\omega) v\right)^n } \right] = E_c^{GW}[G_s] 
\end{align}
and $T_{\rm c}^{\rm RPA} =  E_\text{c}^\text{RPA} -  U_{\rm c}^{\rm RPA}$.
The kinetic energy in sc-RPA is then given by:
\begin{equation}
\label{eq:T-RPA}
    T^{\rm RPA}=T_s+T_{\rm c}^\mathrm{RPA}\, .
\end{equation}
With this reorganization of terms, the kinetic energy in sc-$GW$ can be directly compared
to $T^{\rm RPA}$, and similarly $E_\text{c}^{GW}$ to $U_\text{c}^\text{RPA}$.
 Now the only factor responsible for the difference in these different pairs of terms arises from the difference 
in the input Green functions used to evaluate them.

\begin{figure}
		\epsfig{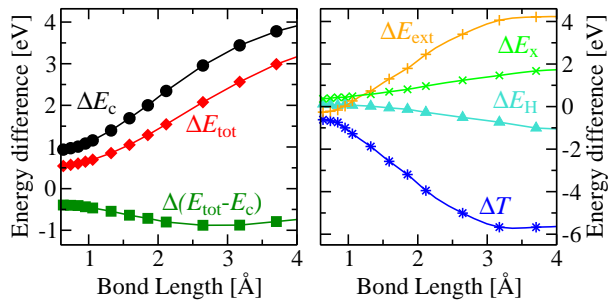}
\caption{\label{fig:gw-rpa} (Color online) Left panel: difference between the sc-$GW$ and sc-RPA total energy ($\Delta E_{\rm tot}$), the correlation energy ($\Delta E_{\rm c}=E_{\rm c}^{GW}-U_{\rm c}^{\rm RPA}$) 
and the remaining terms ($\Delta (E_{\rm tot}-E_{\rm c})$). Right panel: breakdown of the remaining term into the difference of the Hartree ($\Delta E_{\rm H}$), the external ($\Delta E_{\rm ext}$), the exchange ($\Delta E_{\rm x}$), and the kinetic energy ($\Delta T$).}
\end{figure}

We determined the RPA correlation potential following the direct minimization scheme of 
Yang {\it et al.} \cite{Yang2002}. The resulting
orbitals and eigenvalues were used to evaluate
the sc-RPA total energy from Eq.~\ref{Eq:E_Klein-RPA}.
We refer to a previous publication for details of the sc-RPA implementation \cite{Hellgren}.
The sc-$GW$ method -- based on the iterative solution of Eqs.~\ref{Eq:GW_approx} and \ref{Eq:Dyson}
at $\lambda=1$ --
has been implemented in the all-electron localized basis code FHI-aims \cite{blum},
as explained in more detail in Refs.~\cite{caruso/prb/2012,Xinguo/implem_full_author_list}.
The sc-$GW$ total energy was then obtained from Eq.~\ref{eq:gwtot}.

We now turn to an assessment of sc-RPA and sc-$GW$ for  the potential energy curve of H$_2$. In the Supplemental Material, we also show data for other covalently bonded dimers, such as LiH and Li$_2$ \cite{supple_mater}.
In the following, we explicitly refer to non-self-consistent calculations  by appending the suffix @{\it input} to label the Green function used as input. Figure~\ref{fig:h2}  reports the total energy of H$_2$ for different flavors of $GW$ and RPA. For comparison we reproduce the full configuration interaction (CI) potential energy curve of H$_2$ \cite{h2ci1993}, that provides an exact reference for this system. 
We also report the total energy of H$_2$ evaluated from a beyond-$GW$/RPA approach
that incorporates
second-order screened exchange (SOSEX) and renormalized single-excitations in the self-energy \cite{PhysRevLett.106.153003}, referred to in the following as renormalized second order perturbation theory (rPT2) \cite{RPAreview,Xinguo/rpt2}. 
As reported previously~\cite{RPAreview,Fuchs/Gonze/Burke:2005,Henderson/Scuseria:2010,Gorling2011}, non-self-consistent RPA overestimates  the total energy of H$_2$ at the equilibrium bond length. Around the equilibrium distance, the RPA total energy based on exact exchange (OEPx)  and sc-RPA are almost identical and overestimate the total energy by approximately $0.8$ eV, compared to full-CI. At intermediate bond distances and in the dissociation region we see a lowering of the sc-RPA energy compared to RPA@OEPx. 
The spurious ``bump''\cite{Fuchs/Gonze/Burke:2005,RPAreview}, present in all RPA calculations for H$_2$ and other covalently bonded molecules, is reduced in sc-RPA but is still present. The total energy stays below the full-CI energy throughout, indicating a general overestimation of the bonding and dissociation regions.

In agreement with Stan {\it et al.} \cite{stan},  sc-$GW$ provides an accurate total energy  for H$_2$ close to equilibrium. For the Galitskii-Migdal framework, self-consistency is crucial as $G_0W_0$@HF and $G_0W_0$@PBE  largely overestimate the total energy. In contrast, the Klein functional evaluated with the HF Green function (RPA@HF) yields results similar to sc-$GW$.  sc-RPA and sc-$GW$ thus provide a qualitatively similar description  of the energetics of the covalent bond of H$_2$, which results in a slight overestimation of the total energy (see Table \ref{tab:tab}). However, sc-$GW$ is in better agreement with full-CI. Most interestingly, the sc-$GW$ energy is higher than the sc-RPA one. This is in contrast to the exchange only case,  in which the HF 
total energy is always lower than (or equal for a two electron system) the OEPx 
energy \cite{review_OEP}. This is expected, as HF is variational and the local potential in OEPx provides an additional constraint that increases the energy. Conversely, the total energy in  sc-$GW$ has to be higher than in sc-RPA, because the variational procedure yields a maximum at the self-consistent Green function \cite{klein1961,luttingerward1960}.

In the dissociation region,  sc-RPA and sc-$GW$ deviate markedly. 
For sc-RPA the dissociation energy 
is below the full-CI energy and is in rather good agreement with the reference. 
sc-$GW$, on the other hand, fails dramatically in the dissociation limit and with -24.5 eV underestimates the total energy considerably. On the plus side, sc-$GW$ dissociates monotonically and therefore does not show the unphysical ``bump'' present in all RPA-based approaches. Again, both non-self-consistent $G_0W_0$@HF and $G_0W_0$@PBE  energies give better agreement with the reference curve than sc-$GW$.

One could surmise that this qualitatively different behavior originates 
from the different treatment of the kinetic energies that we discussed earlier. 
Figure \ref{fig:gw-rpa}, however, shows that this is not the case. 
At equilibrium the kinetic energy in sc-$GW$ differs only slightly from 
the sc-RPA kinetic energy defined in Eq.~\ref{eq:T-RPA}. This indicates that in the 
bonding regime the AC framework
correctly reproduces the kinetic energy of an interacting system.
At larger bond distances, the kinetic energy differs increasingly in the two approaches.
However, this effect is averaged out
by an opposing change in the external energy that arises from an increasing deviation in the electron densities. 
The same is observed for the Hartree and the exchange energy, although the absolute magnitude of the effect is smaller. 
The total energy difference between sc-$GW$ and sc-RPA can be finally ascribed to
the Coulomb correlation energy 
$\Delta E_{\rm c}=E_{\rm c}^{GW}-U_{\rm c}^{\rm RPA}= E_{\rm c}^{GW}[G]-E_{\rm c}^{GW}[G_s]$, 
as the left panel of Fig.~\ref{fig:gw-rpa} demonstrates.  
Close to equilibrium $\Delta E_{\rm c}$ is of the order of $1$ eV, 
but increases to approximately $4$ eV at larger bond lengths.
This illustrates that it matters decisively whether the correlation energy is evaluated 
with the interacting sc-$GW$ or the non-interacting sc-RPA Green function.

\begin{figure}
\includegraphics[width=0.48\textwidth]{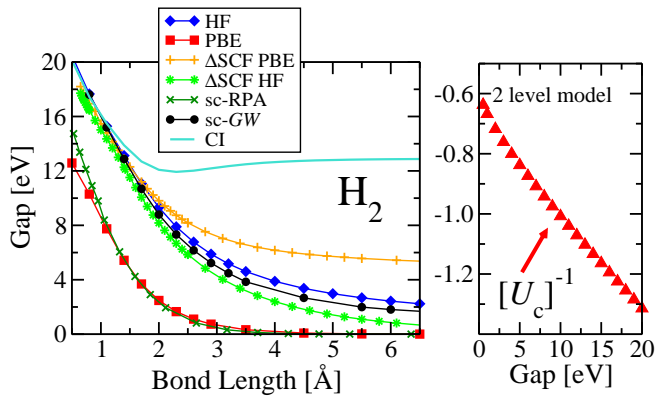}
\caption{\label{fig:gw-rpa_gap} (Color online) 
Left panel: HOMO-LUMO gap extracted from the
sc-$GW$ spectral function and from 
HF, PBE, and sc-RPA eigenvalues. The HOMO-LUMO gap evaluated from PBE and HF total energy differences ($\Delta$SCF)
is included for comparison. Full configuration interaction (CI) calculation were done with a aug-cc-pVDZ basis set.
Right panel: Inverse of the RPA Coulomb correlation energy $U_{\rm c}^{\rm RPA}$ 
for a two-level model as a function of the HOMO-LUMO gap.}
\end{figure}

%
Why the difference is so pronounced at dissociation is still an open question. 
A potential explanation can be found in the 
inverse dependence of the RPA Coulomb correlation energy on the gap 
between the highest occupied and the lowest unoccupied molecular 
orbital (HOMO and LUMO, respectively). This is exemplified by the 
right panel of Fig.~\ref{fig:gw-rpa_gap}, which shows the inverse of $U_{\rm c}$ as a function 
of the gap for a simplified two level system. The large value of $U_{\rm c}$ 
obtained from sc-RPA for H$_2$ at dissociation can therefore be traced 
back to the small HOMO-LUMO gap (left panel of Fig. 4) of the RPA 
Green function,  as also illustrated in the Supplemental Material of Ref.~\onlinecite{Gorling2011}. 
In contrast, due to the spatial non-locality of the self-energy,
the HOMO-LUMO gap of the HF and $GW$ Green functions 
is much larger at every given bond distance. 
This leads in turn to a smaller Coulomb correlation energy 
for sc-$GW$ and HF-based perturbative methods.

In conclusion, 
we have compared MBPT in the $GW$ approximation to DFT in the RPA.
We found that the density functional description is superior at dissociation, 
yielding a total energy in qualitative agreement with the exact energy along 
the entire dissociation curve. 
These results illustrate how MBPT and DFT based approaches deal with multi-reference ground-states. 
We demonstrated that in a DFT-based 
framework the closure of the (KS) HOMO-LUMO gap
is in part responsible for the 
improved description at dissociation,
i.e., static correlation is better 
accounted for in sc-RPA, than in sc-$GW$.
The same effect in Green function theory has to 
be achieved by the right (potentially infinite) 
set of diagrams. 
We conclude that static and local approximations of exchange-correlation potentials 
-- as opposed to non-local,  frequency dependent self-energy approximations --
are more effective in describing the dissociation regime 
of covalently bonded molecules.

\acknowledgements
DRR gratefully acknowledges financial support from the German National Academy of Sciences -- Leopoldina under grant number LPDS 2011-15.
AR  acknowledges  support from the European Research Council (ERC-2010- AdG -267374), Spanish Grant (FIS2011- 65702-C02-01 and PIB2010US-00652), Grupos Consolidados del Gobierno Vasco (IT-319-07), and EC project CRONOS (280879-2).


\begin{thebibliography}{39}%
\makeatletter
\providecommand \@ifxundefined [1]{%
 \@ifx{#1\undefined}
}%
\providecommand \@ifnum [1]{%
 \ifnum #1\expandafter \@firstoftwo
 \else \expandafter \@secondoftwo
 \fi
}%
\providecommand \@ifx [1]{%
 \ifx #1\expandafter \@firstoftwo
 \else \expandafter \@secondoftwo
 \fi
}%
\providecommand \natexlab [1]{#1}%
\providecommand \enquote  [1]{``#1''}%
\providecommand \bibnamefont  [1]{#1}%
\providecommand \bibfnamefont [1]{#1}%
\providecommand \citenamefont [1]{#1}%
\providecommand \href@noop [0]{\@secondoftwo}%
\providecommand \href [0]{\begingroup \@sanitize@url \@href}%
\providecommand \@href[1]{\@@startlink{#1}\@@href}%
\providecommand \@@href[1]{\endgroup#1\@@endlink}%
\providecommand \@sanitize@url [0]{\catcode `\\12\catcode `\$12\catcode
  `\&12\catcode `\#12\catcode `\^12\catcode `\_12\catcode `\%12\relax}%
\providecommand \@@startlink[1]{}%
\providecommand \@@endlink[0]{}%
\providecommand \url  [0]{\begingroup\@sanitize@url \@url }%
\providecommand \@url [1]{\endgroup\@href {#1}{\urlprefix }}%
\providecommand \urlprefix  [0]{URL }%
\providecommand \Eprint [0]{\href }%
\providecommand \doibase [0]{http://dx.doi.org/}%
\providecommand \selectlanguage [0]{\@gobble}%
\providecommand \bibinfo  [0]{\@secondoftwo}%
\providecommand \bibfield  [0]{\@secondoftwo}%
\providecommand \translation [1]{[#1]}%
\providecommand \BibitemOpen [0]{}%
\providecommand \bibitemStop [0]{}%
\providecommand \bibitemNoStop [0]{.\EOS\space}%
\providecommand \EOS [0]{\spacefactor3000\relax}%
\providecommand \BibitemShut  [1]{\csname bibitem#1\endcsname}%
\let\auto@bib@innerbib\@empty
\bibitem [{\citenamefont {Hohenberg}\ and\ \citenamefont
  {Kohn}(1964)}]{hohenbergkohn}%
  \BibitemOpen
  \bibfield  {author} {\bibinfo {author} {\bibfnamefont {P.}~\bibnamefont
  {Hohenberg}}\ and\ \bibinfo {author} {\bibfnamefont {W.}~\bibnamefont
  {Kohn}},\ }\href {\doibase 10.1103/PhysRev.136.B864} {\bibfield  {journal}
  {\bibinfo  {journal} {Phys. Rev.}\ }\textbf {\bibinfo {volume} {136}},\
  \bibinfo {pages} {B864} (\bibinfo {year} {1964})}\BibitemShut {NoStop}%
\bibitem [{\citenamefont {Kohn}\ and\ \citenamefont
  {Sham}(1965)}]{kohnsham1965}%
  \BibitemOpen
  \bibfield  {author} {\bibinfo {author} {\bibfnamefont {W.}~\bibnamefont
  {Kohn}}\ and\ \bibinfo {author} {\bibfnamefont {L.~J.}\ \bibnamefont
  {Sham}},\ }\href {\doibase 10.1103/PhysRev.140.A1133} {\bibfield  {journal}
  {\bibinfo  {journal} {Phys. Rev.}\ }\textbf {\bibinfo {volume} {140}},\
  \bibinfo {pages} {A1133} (\bibinfo {year} {1965})}\BibitemShut {NoStop}%
\bibitem [{\citenamefont {Dreizler}\ and\ \citenamefont
  {Gross}(1990)}]{Dreizler}%
  \BibitemOpen
  \bibfield  {author} {\bibinfo {author} {\bibfnamefont {R.~M.}\ \bibnamefont
  {Dreizler}}\ and\ \bibinfo {author} {\bibfnamefont {E.~K.~U.}\ \bibnamefont
  {Gross}},\ }\href@noop {} {\emph {\bibinfo {title} {Density Functional
  Theory}}}\ (\bibinfo  {publisher} {Springer, Berlin},\ \bibinfo {year}
  {1990})\BibitemShut {NoStop}%
\bibitem [{\citenamefont {Szabo}\ and\ \citenamefont
  {Ostlund}(1989)}]{Szabo/Ostlund:1989}%
  \BibitemOpen
  \bibfield  {author} {\bibinfo {author} {\bibfnamefont {A.}~\bibnamefont
  {Szabo}}\ and\ \bibinfo {author} {\bibfnamefont {N.~S.}\ \bibnamefont
  {Ostlund}},\ }\href@noop {} {\emph {\bibinfo {title} {\textit{Modern Quantum
  Chemistry: Introduction to Advanced Electronic Structure Theory}}}}\
  (\bibinfo  {publisher} {McGraw-Hill},\ \bibinfo {address} {New York},\
  \bibinfo {year} {1989})\BibitemShut {NoStop}%
\bibitem [{\citenamefont {Fetter}\ and\ \citenamefont
  {Walecka}(2003)}]{fetter}%
  \BibitemOpen
  \bibfield  {author} {\bibinfo {author} {\bibfnamefont {A.~L.}\ \bibnamefont
  {Fetter}}\ and\ \bibinfo {author} {\bibfnamefont {J.~D.}\ \bibnamefont
  {Walecka}},\ }\href@noop {} {\emph {\bibinfo {title} {Quantum Theory of
  {Many-Particle} Systems}}}\ (\bibinfo  {publisher} {Dover Publications},\
  \bibinfo {year} {2003})\BibitemShut {NoStop}%
\bibitem [{\citenamefont {Hedin}(1965)}]{Hedin1965}%
  \BibitemOpen
  \bibfield  {author} {\bibinfo {author} {\bibfnamefont {L.}~\bibnamefont
  {Hedin}},\ }\href {\doibase 10.1103/PhysRev.139.A796} {\bibfield  {journal}
  {\bibinfo  {journal} {Phys. Rev.}\ }\textbf {\bibinfo {volume} {139}},\
  \bibinfo {pages} {A796} (\bibinfo {year} {1965})}\BibitemShut {NoStop}%
\bibitem [{\citenamefont {Bartlett}\ and\ \citenamefont
  {Musia\l{}}(2007)}]{RevModPhys.79.291}%
  \BibitemOpen
  \bibfield  {author} {\bibinfo {author} {\bibfnamefont {R.~J.}\ \bibnamefont
  {Bartlett}}\ and\ \bibinfo {author} {\bibfnamefont {M.}~\bibnamefont
  {Musia\l{}}},\ }\href {\doibase 10.1103/RevModPhys.79.291} {\bibfield
  {journal} {\bibinfo  {journal} {Rev. Mod. Phys.}\ }\textbf {\bibinfo {volume}
  {79}},\ \bibinfo {pages} {291} (\bibinfo {year} {2007})}\BibitemShut
  {NoStop}%
\bibitem [{\citenamefont {Foulkes}\ \emph {et~al.}(2001)\citenamefont
  {Foulkes}, \citenamefont {Mitas}, \citenamefont {Needs},\ and\ \citenamefont
  {Rajagopal}}]{RevModPhys.73.33}%
  \BibitemOpen
  \bibfield  {author} {\bibinfo {author} {\bibfnamefont {W.~M.~C.}\
  \bibnamefont {Foulkes}}, \bibinfo {author} {\bibfnamefont {L.}~\bibnamefont
  {Mitas}}, \bibinfo {author} {\bibfnamefont {R.~J.}\ \bibnamefont {Needs}}, \
  and\ \bibinfo {author} {\bibfnamefont {G.}~\bibnamefont {Rajagopal}},\ }\href
  {\doibase 10.1103/RevModPhys.73.33} {\bibfield  {journal} {\bibinfo
  {journal} {Rev. Mod. Phys.}\ }\textbf {\bibinfo {volume} {73}},\ \bibinfo
  {pages} {33} (\bibinfo {year} {2001})}\BibitemShut {NoStop}%
\bibitem [{\citenamefont {Baym}\ and\ \citenamefont
  {Kadanoff}(1961)}]{baymkadanoff}%
  \BibitemOpen
  \bibfield  {author} {\bibinfo {author} {\bibfnamefont {G.}~\bibnamefont
  {Baym}}\ and\ \bibinfo {author} {\bibfnamefont {L.~P.}\ \bibnamefont
  {Kadanoff}},\ }\href {\doibase 10.1103/PhysRev.124.287} {\bibfield  {journal}
  {\bibinfo  {journal} {Phys. Rev.}\ }\textbf {\bibinfo {volume} {124}},\
  \bibinfo {pages} {287} (\bibinfo {year} {1961})}\BibitemShut {NoStop}%
\bibitem [{\citenamefont {Luttinger}\ and\ \citenamefont
  {Ward}(1960)}]{luttingerward1960}%
  \BibitemOpen
  \bibfield  {author} {\bibinfo {author} {\bibfnamefont {J.~M.}\ \bibnamefont
  {Luttinger}}\ and\ \bibinfo {author} {\bibfnamefont {J.~C.}\ \bibnamefont
  {Ward}},\ }\href {\doibase 10.1103/PhysRev.118.1417} {\bibfield  {journal}
  {\bibinfo  {journal} {Phys. Rev.}\ }\textbf {\bibinfo {volume} {118}},\
  \bibinfo {pages} {1417} (\bibinfo {year} {1960})}\BibitemShut {NoStop}%
\bibitem [{\citenamefont {G\"orling}\ and\ \citenamefont
  {Ernzerhof}(1995)}]{Gorling/Ernzerhof/OEP/HF}%
  \BibitemOpen
  \bibfield  {author} {\bibinfo {author} {\bibfnamefont {A.}~\bibnamefont
  {G\"orling}}\ and\ \bibinfo {author} {\bibfnamefont {M.}~\bibnamefont
  {Ernzerhof}},\ }\href {\doibase 10.1103/PhysRevA.51.4501} {\bibfield
  {journal} {\bibinfo  {journal} {Phys. Rev. A}\ }\textbf {\bibinfo {volume}
  {51}},\ \bibinfo {pages} {4501} (\bibinfo {year} {1995})}\BibitemShut
  {NoStop}%
\bibitem [{\citenamefont {K\"ummel}\ and\ \citenamefont
  {Kronik}(2008)}]{review_OEP}%
  \BibitemOpen
  \bibfield  {author} {\bibinfo {author} {\bibfnamefont {S.}~\bibnamefont
  {K\"ummel}}\ and\ \bibinfo {author} {\bibfnamefont {L.}~\bibnamefont
  {Kronik}},\ }\href {\doibase 10.1103/RevModPhys.80.3} {\bibfield  {journal}
  {\bibinfo  {journal} {Rev. Mod. Phys.}\ }\textbf {\bibinfo {volume} {80}},\
  \bibinfo {pages} {3} (\bibinfo {year} {2008})}\BibitemShut {NoStop}%
\bibitem [{\citenamefont {Bohm}\ and\ \citenamefont
  {Pines}(1953)}]{Bohm/Pines:1953}%
  \BibitemOpen
  \bibfield  {author} {\bibinfo {author} {\bibfnamefont {D.}~\bibnamefont
  {Bohm}}\ and\ \bibinfo {author} {\bibfnamefont {D.}~\bibnamefont {Pines}},\
  }\href@noop {} {\bibfield  {journal} {\bibinfo  {journal} {Phys. Rev.}\
  }\textbf {\bibinfo {volume} {92}},\ \bibinfo {pages} {609} (\bibinfo {year}
  {1953})}\BibitemShut {NoStop}%
\bibitem [{\citenamefont {Gell-Mann}\ and\ \citenamefont
  {Brueckner}(1957)}]{Gell-Mann/Brueckner:1957}%
  \BibitemOpen
  \bibfield  {author} {\bibinfo {author} {\bibfnamefont {M.}~\bibnamefont
  {Gell-Mann}}\ and\ \bibinfo {author} {\bibfnamefont {K.~A.}\ \bibnamefont
  {Brueckner}},\ }\href@noop {} {\bibfield  {journal} {\bibinfo  {journal}
  {Phys. Rev.}\ }\textbf {\bibinfo {volume} {106}},\ \bibinfo {pages} {364}
  (\bibinfo {year} {1957})}\BibitemShut {NoStop}%
\bibitem [{\citenamefont {Langreth}\ and\ \citenamefont
  {Perdew}(1977)}]{Langreth/Perdew:1977}%
  \BibitemOpen
  \bibfield  {author} {\bibinfo {author} {\bibfnamefont {D.~C.}\ \bibnamefont
  {Langreth}}\ and\ \bibinfo {author} {\bibfnamefont {J.~P.}\ \bibnamefont
  {Perdew}},\ }\href@noop {} {\bibfield  {journal} {\bibinfo  {journal} {Phys.
  Rev. B}\ }\textbf {\bibinfo {volume} {15}},\ \bibinfo {pages} {2884}
  (\bibinfo {year} {1977})}\BibitemShut {NoStop}%
\bibitem [{\citenamefont {Ren}\ \emph {et~al.}(2012{\natexlab{a}})\citenamefont
  {Ren}, \citenamefont {Rinke}, \citenamefont {Joas},\ and\ \citenamefont
  {Scheffler}}]{RPAreview}%
  \BibitemOpen
  \bibfield  {author} {\bibinfo {author} {\bibfnamefont {X.}~\bibnamefont
  {Ren}}, \bibinfo {author} {\bibfnamefont {P.}~\bibnamefont {Rinke}}, \bibinfo
  {author} {\bibfnamefont {C.}~\bibnamefont {Joas}}, \ and\ \bibinfo {author}
  {\bibfnamefont {M.}~\bibnamefont {Scheffler}},\ }\href@noop {} {\bibfield
  {journal} {\bibinfo  {journal} {J. Mater. Sci.}\ }\textbf {\bibinfo {volume}
  {47}},\ \bibinfo {pages} {21} (\bibinfo {year}
  {2012}{\natexlab{a}})}\BibitemShut {NoStop}%
\bibitem [{\citenamefont {Aryasetiawan}\ and\ \citenamefont
  {Gunnarsson}(1998)}]{thegwmethod}%
  \BibitemOpen
  \bibfield  {author} {\bibinfo {author} {\bibfnamefont {F.}~\bibnamefont
  {Aryasetiawan}}\ and\ \bibinfo {author} {\bibfnamefont {O.}~\bibnamefont
  {Gunnarsson}},\ }\href {http://stacks.iop.org/0034-4885/61/i=3/a=002}
  {\bibfield  {journal} {\bibinfo  {journal} {Rep. Prog. Phys.}\ }\textbf
  {\bibinfo {volume} {61}},\ \bibinfo {pages} {237} (\bibinfo {year}
  {1998})}\BibitemShut {NoStop}%
\bibitem [{\citenamefont {Wolniewicz}(1993)}]{h2ci1993}%
  \BibitemOpen
  \bibfield  {author} {\bibinfo {author} {\bibfnamefont {L.}~\bibnamefont
  {Wolniewicz}},\ }\href {\doibase DOI:10.1063/1.465303} {\bibfield  {journal}
  {\bibinfo  {journal} {J. Chem. Phys.}\ }\textbf {\bibinfo {volume} {99}},\
  \bibinfo {pages} {1851} (\bibinfo {year} {1993})}\BibitemShut {NoStop}%
\bibitem [{\citenamefont {Thom H.~Dunning}(1989)}]{gaussianbasis1989}%
  \BibitemOpen
  \bibfield  {author} {\bibinfo {author} {\bibfnamefont {J.}~\bibnamefont {Thom
  H.~Dunning}},\ }\href {\doibase 10.1063/1.456153} {\bibfield  {journal}
  {\bibinfo  {journal} {J. Chem. Phys.}\ }\textbf {\bibinfo {volume} {90}},\
  \bibinfo {pages} {1007} (\bibinfo {year} {1989})}\BibitemShut {NoStop}%
\bibitem [{\citenamefont {Niquet}\ \emph {et~al.}(2003)\citenamefont {Niquet},
  \citenamefont {Fuchs},\ and\ \citenamefont {Gonze}}]{PhysRevA.68.032507}%
  \BibitemOpen
  \bibfield  {author} {\bibinfo {author} {\bibfnamefont {Y.~M.}\ \bibnamefont
  {Niquet}}, \bibinfo {author} {\bibfnamefont {M.}~\bibnamefont {Fuchs}}, \
  and\ \bibinfo {author} {\bibfnamefont {X.}~\bibnamefont {Gonze}},\ }\href
  {\doibase 10.1103/PhysRevA.68.032507} {\bibfield  {journal} {\bibinfo
  {journal} {Phys. Rev. A}\ }\textbf {\bibinfo {volume} {68}},\ \bibinfo
  {pages} {032507} (\bibinfo {year} {2003})}\BibitemShut {NoStop}%
\bibitem [{sup()}]{supple_mater}%
  \BibitemOpen
  \href@noop {} {}\bibinfo {note} {See the supplemental material at
  \url{address}}\BibitemShut {NoStop}%
\bibitem [{\citenamefont {Sham}\ and\ \citenamefont
  {Schl\"uter}(1983)}]{ShamSchluter}%
  \BibitemOpen
  \bibfield  {author} {\bibinfo {author} {\bibfnamefont {L.~J.}\ \bibnamefont
  {Sham}}\ and\ \bibinfo {author} {\bibfnamefont {M.}~\bibnamefont
  {Schl\"uter}},\ }\href {\doibase 10.1103/PhysRevLett.51.1888} {\bibfield
  {journal} {\bibinfo  {journal} {Phys. Rev. Lett.}\ }\textbf {\bibinfo
  {volume} {51}},\ \bibinfo {pages} {1888} (\bibinfo {year}
  {1983})}\BibitemShut {NoStop}%
\bibitem [{\citenamefont {Casida}(1995)}]{Casida:1995}%
  \BibitemOpen
  \bibfield  {author} {\bibinfo {author} {\bibfnamefont {M.~E.}\ \bibnamefont
  {Casida}},\ }\href@noop {} {\bibfield  {journal} {\bibinfo  {journal} {Phys.\
  Rev.\ A}\ }\textbf {\bibinfo {volume} {51}},\ \bibinfo {pages} {2005}
  (\bibinfo {year} {1995})}\BibitemShut {NoStop}%
\bibitem [{\citenamefont {Klein}(1961)}]{klein1961}%
  \BibitemOpen
  \bibfield  {author} {\bibinfo {author} {\bibfnamefont {A.}~\bibnamefont
  {Klein}},\ }\href {\doibase 10.1103/PhysRev.121.950} {\bibfield  {journal}
  {\bibinfo  {journal} {Phys. Rev.}\ }\textbf {\bibinfo {volume} {121}},\
  \bibinfo {pages} {950} (\bibinfo {year} {1961})}\BibitemShut {NoStop}%
\bibitem [{\citenamefont {Dahlen}\ \emph {et~al.}(2006)\citenamefont {Dahlen},
  \citenamefont {van Leeuwen},\ and\ \citenamefont {von
  Barth}}]{dahlenleeuwen2006}%
  \BibitemOpen
  \bibfield  {author} {\bibinfo {author} {\bibfnamefont {N.~E.}\ \bibnamefont
  {Dahlen}}, \bibinfo {author} {\bibfnamefont {R.}~\bibnamefont {van Leeuwen}},
  \ and\ \bibinfo {author} {\bibfnamefont {U.}~\bibnamefont {von Barth}},\
  }\href {\doibase 10.1103/PhysRevA.73.012511} {\bibfield  {journal} {\bibinfo
  {journal} {Phys. Rev. A}\ }\textbf {\bibinfo {volume} {73}},\ \bibinfo
  {pages} {012511} (\bibinfo {year} {2006})}\BibitemShut {NoStop}%
\bibitem [{\citenamefont {Holm}\ and\ \citenamefont
  {Aryasetiawan}(2000)}]{galitskiimigdal}%
  \BibitemOpen
  \bibfield  {author} {\bibinfo {author} {\bibfnamefont {B.}~\bibnamefont
  {Holm}}\ and\ \bibinfo {author} {\bibfnamefont {F.}~\bibnamefont
  {Aryasetiawan}},\ }\href {\doibase 10.1103/PhysRevB.62.4858} {\bibfield
  {journal} {\bibinfo  {journal} {Phys. Rev. B}\ }\textbf {\bibinfo {volume}
  {62}},\ \bibinfo {pages} {4858} (\bibinfo {year} {2000})}\BibitemShut
  {NoStop}%
\bibitem [{\citenamefont {Baym}(1962)}]{baym}%
  \BibitemOpen
  \bibfield  {author} {\bibinfo {author} {\bibfnamefont {G.}~\bibnamefont
  {Baym}},\ }\href {\doibase 10.1103/PhysRev.127.1391} {\bibfield  {journal}
  {\bibinfo  {journal} {Phys. Rev.}\ }\textbf {\bibinfo {volume} {127}},\
  \bibinfo {pages} {1391} (\bibinfo {year} {1962})}\BibitemShut {NoStop}%
\bibitem [{\citenamefont {Dahlen}\ and\ \citenamefont {van
  Leeuwen}(2005)}]{dahlenleeuwen2005}%
  \BibitemOpen
  \bibfield  {author} {\bibinfo {author} {\bibfnamefont {N.~E.}\ \bibnamefont
  {Dahlen}}\ and\ \bibinfo {author} {\bibfnamefont {R.}~\bibnamefont {van
  Leeuwen}},\ }\href {\doibase 10.1063/1.1884965} {\bibfield  {journal}
  {\bibinfo  {journal} {J. Chem. Phys.}\ }\textbf {\bibinfo {volume} {122}},\
  \bibinfo {eid} {164102} (\bibinfo {year} {2005})}\BibitemShut {NoStop}%
\bibitem [{\citenamefont {Caruso}\ \emph {et~al.}(2012)\citenamefont {Caruso},
  \citenamefont {Rinke}, \citenamefont {Ren}, \citenamefont {Scheffler},\ and\
  \citenamefont {Rubio}}]{caruso/prb/2012}%
  \BibitemOpen
  \bibfield  {author} {\bibinfo {author} {\bibfnamefont {F.}~\bibnamefont
  {Caruso}}, \bibinfo {author} {\bibfnamefont {P.}~\bibnamefont {Rinke}},
  \bibinfo {author} {\bibfnamefont {X.}~\bibnamefont {Ren}}, \bibinfo {author}
  {\bibfnamefont {M.}~\bibnamefont {Scheffler}}, \ and\ \bibinfo {author}
  {\bibfnamefont {A.}~\bibnamefont {Rubio}},\ }\href {\doibase
  10.1103/PhysRevB.86.081102} {\bibfield  {journal} {\bibinfo  {journal} {Phys.
  Rev. B}\ }\textbf {\bibinfo {volume} {86}},\ \bibinfo {pages} {081102(R)}
  (\bibinfo {year} {2012})}\BibitemShut {NoStop}%
\bibitem [{\citenamefont {Yang}\ and\ \citenamefont {Wu}(2002)}]{Yang2002}%
  \BibitemOpen
  \bibfield  {author} {\bibinfo {author} {\bibfnamefont {W.}~\bibnamefont
  {Yang}}\ and\ \bibinfo {author} {\bibfnamefont {Q.}~\bibnamefont {Wu}},\
  }\href {\doibase 10.1103/PhysRevLett.89.143002} {\bibfield  {journal}
  {\bibinfo  {journal} {Phys. Rev. Lett.}\ }\textbf {\bibinfo {volume} {89}},\
  \bibinfo {pages} {143002} (\bibinfo {year} {2002})}\BibitemShut {NoStop}%
\bibitem [{\citenamefont {Hellgren}\ \emph {et~al.}(2012)\citenamefont
  {Hellgren}, \citenamefont {Rohr},\ and\ \citenamefont {Gross}}]{Hellgren}%
  \BibitemOpen
  \bibfield  {author} {\bibinfo {author} {\bibfnamefont {M.}~\bibnamefont
  {Hellgren}}, \bibinfo {author} {\bibfnamefont {D.~R.}\ \bibnamefont {Rohr}},
  \ and\ \bibinfo {author} {\bibfnamefont {E.~K.~U.}\ \bibnamefont {Gross}},\
  }\href {\doibase 10.1063/1.3676174} {\bibfield  {journal} {\bibinfo
  {journal} {J. Chem. Phys.}\ }\textbf {\bibinfo {volume} {136}},\ \bibinfo
  {eid} {034106} (\bibinfo {year} {2012})}\BibitemShut {NoStop}%
\bibitem [{\citenamefont {Blum}\ \emph {et~al.}(2009)\citenamefont {Blum},
  \citenamefont {Gehrke}, \citenamefont {Hanke}, \citenamefont {Havu},
  \citenamefont {Havu}, \citenamefont {Ren}, \citenamefont {Reuter},\ and\
  \citenamefont {Scheffler}}]{blum}%
  \BibitemOpen
  \bibfield  {author} {\bibinfo {author} {\bibfnamefont {V.}~\bibnamefont
  {Blum}}, \bibinfo {author} {\bibfnamefont {R.}~\bibnamefont {Gehrke}},
  \bibinfo {author} {\bibfnamefont {F.}~\bibnamefont {Hanke}}, \bibinfo
  {author} {\bibfnamefont {P.}~\bibnamefont {Havu}}, \bibinfo {author}
  {\bibfnamefont {V.}~\bibnamefont {Havu}}, \bibinfo {author} {\bibfnamefont
  {X.}~\bibnamefont {Ren}}, \bibinfo {author} {\bibfnamefont {K.}~\bibnamefont
  {Reuter}}, \ and\ \bibinfo {author} {\bibfnamefont {M.}~\bibnamefont
  {Scheffler}},\ }\href {\doibase DOI: 10.1016/j.cpc.2009.06.022} {\bibfield
  {journal} {\bibinfo  {journal} {Comp. Phys. Comm.}\ }\textbf {\bibinfo
  {volume} {180}},\ \bibinfo {pages} {2175 } (\bibinfo {year}
  {2009})}\BibitemShut {NoStop}%
\bibitem [{\citenamefont {Ren}\ \emph {et~al.}(2012{\natexlab{b}})\citenamefont
  {Ren}, \citenamefont {Rinke}, \citenamefont {Blum}, \citenamefont
  {Wieferink}, \citenamefont {Tkatchenko}, \citenamefont {Andrea},
  \citenamefont {Reuter}, \citenamefont {Blum},\ and\ \citenamefont
  {Scheffler}}]{Xinguo/implem_full_author_list}%
  \BibitemOpen
  \bibfield  {author} {\bibinfo {author} {\bibfnamefont {X.}~\bibnamefont
  {Ren}}, \bibinfo {author} {\bibfnamefont {P.}~\bibnamefont {Rinke}}, \bibinfo
  {author} {\bibfnamefont {V.}~\bibnamefont {Blum}}, \bibinfo {author}
  {\bibfnamefont {J.}~\bibnamefont {Wieferink}}, \bibinfo {author}
  {\bibfnamefont {A.}~\bibnamefont {Tkatchenko}}, \bibinfo {author}
  {\bibfnamefont {S.}~\bibnamefont {Andrea}}, \bibinfo {author} {\bibfnamefont
  {K.}~\bibnamefont {Reuter}}, \bibinfo {author} {\bibfnamefont
  {V.}~\bibnamefont {Blum}}, \ and\ \bibinfo {author} {\bibfnamefont
  {M.}~\bibnamefont {Scheffler}},\ }\href@noop {} {\bibfield  {journal}
  {\bibinfo  {journal} {New J. Phys.}\ }\textbf {\bibinfo {volume} {14}},\
  \bibinfo {pages} {053020} (\bibinfo {year} {2012}{\natexlab{b}})}\BibitemShut
  {NoStop}%
\bibitem [{\citenamefont {Ren}\ \emph {et~al.}(2011)\citenamefont {Ren},
  \citenamefont {Tkatchenko}, \citenamefont {Rinke},\ and\ \citenamefont
  {Scheffler}}]{PhysRevLett.106.153003}%
  \BibitemOpen
  \bibfield  {author} {\bibinfo {author} {\bibfnamefont {X.}~\bibnamefont
  {Ren}}, \bibinfo {author} {\bibfnamefont {A.}~\bibnamefont {Tkatchenko}},
  \bibinfo {author} {\bibfnamefont {P.}~\bibnamefont {Rinke}}, \ and\ \bibinfo
  {author} {\bibfnamefont {M.}~\bibnamefont {Scheffler}},\ }\href {\doibase
  10.1103/PhysRevLett.106.153003} {\bibfield  {journal} {\bibinfo  {journal}
  {Phys. Rev. Lett.}\ }\textbf {\bibinfo {volume} {106}},\ \bibinfo {pages}
  {153003} (\bibinfo {year} {2011})}\BibitemShut {NoStop}%
\bibitem [{\citenamefont {Ren}\ \emph {et~al.}()\citenamefont {Ren},
  \citenamefont {Rinke}, \citenamefont {Scuseria},\ and\ \citenamefont
  {Scheffler}}]{Xinguo/rpt2}%
  \BibitemOpen
  \bibfield  {author} {\bibinfo {author} {\bibfnamefont {X.}~\bibnamefont
  {Ren}}, \bibinfo {author} {\bibfnamefont {P.}~\bibnamefont {Rinke}}, \bibinfo
  {author} {\bibfnamefont {G.~E.}\ \bibnamefont {Scuseria}}, \ and\ \bibinfo
  {author} {\bibfnamefont {M.}~\bibnamefont {Scheffler}},\ }\href@noop {}
  {\bibinfo  {journal} {in preparation}\ }\BibitemShut {NoStop}%
\bibitem [{\citenamefont {Fuchs}\ \emph {et~al.}(2005)\citenamefont {Fuchs},
  \citenamefont {Niquet}, \citenamefont {Gonze},\ and\ \citenamefont
  {Burke}}]{Fuchs/Gonze/Burke:2005}%
  \BibitemOpen
\bibfield  {journal} {  }\bibfield  {author} {\bibinfo {author} {\bibfnamefont
  {M.}~\bibnamefont {Fuchs}}, \bibinfo {author} {\bibfnamefont {Y.-M.}\
  \bibnamefont {Niquet}}, \bibinfo {author} {\bibfnamefont {X.}~\bibnamefont
  {Gonze}}, \ and\ \bibinfo {author} {\bibfnamefont {K.}~\bibnamefont
  {Burke}},\ }\href@noop {} {\bibfield  {journal} {\bibinfo  {journal} {J.
  Chem. Phys.}\ }\textbf {\bibinfo {volume} {122}},\ \bibinfo {pages} {094116}
  (\bibinfo {year} {2005})}\BibitemShut {NoStop}%
\bibitem [{\citenamefont {Henderson}\ and\ \citenamefont
  {Scuseria}(2010)}]{Henderson/Scuseria:2010}%
  \BibitemOpen
  \bibfield  {author} {\bibinfo {author} {\bibfnamefont {T.~M.}\ \bibnamefont
  {Henderson}}\ and\ \bibinfo {author} {\bibfnamefont {G.~E.}\ \bibnamefont
  {Scuseria}},\ }\href@noop {} {\bibfield  {journal} {\bibinfo  {journal} {Mol.
  Phys.}\ }\textbf {\bibinfo {volume} {108}},\ \bibinfo {pages} {2511}
  (\bibinfo {year} {2010})}\BibitemShut {NoStop}%
\bibitem [{\citenamefont {Hesselmann}\ and\ \citenamefont
  {G\"orling}(2011)}]{Gorling2011}%
  \BibitemOpen
  \bibfield  {author} {\bibinfo {author} {\bibfnamefont {A.}~\bibnamefont
  {Hesselmann}}\ and\ \bibinfo {author} {\bibfnamefont {A.}~\bibnamefont
  {G\"orling}},\ }\href {\doibase 10.1103/PhysRevLett.106.093001} {\bibfield
  {journal} {\bibinfo  {journal} {Phys. Rev. Lett.}\ }\textbf {\bibinfo
  {volume} {106}},\ \bibinfo {pages} {093001} (\bibinfo {year}
  {2011})}\BibitemShut {NoStop}%
\bibitem [{\citenamefont {Stan}\ \emph {et~al.}(2009)\citenamefont {Stan},
  \citenamefont {Dahlen},\ and\ \citenamefont {van Leeuwen}}]{stan}%
  \BibitemOpen
  \bibfield  {author} {\bibinfo {author} {\bibfnamefont {A.}~\bibnamefont
  {Stan}}, \bibinfo {author} {\bibfnamefont {N.~E.}\ \bibnamefont {Dahlen}}, \
  and\ \bibinfo {author} {\bibfnamefont {R.}~\bibnamefont {van Leeuwen}},\
  }\href {\doibase 10.1063/1.3089567} {\bibfield  {journal} {\bibinfo
  {journal} {J. Chem. Phys.}\ }\textbf {\bibinfo {volume} {130}},\ \bibinfo
  {eid} {114105} (\bibinfo {year} {2009})}\BibitemShut {NoStop}%
\end{thebibliography}

%

\end{document}